\documentclass{aastex}

\newcommand{\kms}{\mbox{km s$^{-1}$}}

\newcommand{\Msun}{\mbox{\,M$_{\sun}$}}
\newcommand{\Lsun}{\mbox{L$_{\sun}~$}}
\newcommand{\um}{\mbox{$\mu$m }} 

\newcommand{\skipthis}[1]{}
\newcommand{\hii}{\mbox{\ion{H}{2}}}

\shortauthors{Guzm{\'a}n et al.}
\shorttitle{A collimated jet toward IRAS 16562-3959}
\usepackage{amssymb,amsmath,natbib,graphicx}
\begin{document}


\title{A string of radio emission associated with IRAS 16562-3959: a 
collimated jet emanating from a luminous massive YSO}

\author{Andres Guzm{\'{a}}n\altaffilmark{1}, Guido Garay\altaffilmark{1},
\and Kate J. Brooks\altaffilmark{2}}


\altaffiltext{1}{Departamento de Astronom\'{\i}a, Universidad de Chile, Camino el
  Observatorio 1515, Las Condes, Santiago, Chile}\altaffiltext{2}{
CSIRO Astronomy and Space Science, P.O. Box 76, Epping 1710 NSW,
  Australia}

\begin{abstract}

We report the discovery made using the Australia Telescope Compact Array of a remarkable
string of radio emission towards IRAS 16562-3959, a luminous infrared source with a
bolometric luminosity of $7.0\times10^4$ \Lsun.  The radio emission arises from a compact,
bright central component, two inner lobes, which are separated by about 7\arcsec\ and
symmetrically offset from the central source, and two outer lobes which are separated by
about 45\arcsec.  The emission from the central object has a spectral index between 1.4
and 8.6 GHz of $0.85\pm0.15$, consistent with free-free emission from a thermal jet.  The
radio emission from the lobes have spectral indices in the range characteristic of thermal
emission. We suggest that the emission from the lobes arises in shocks resulting from the
interaction of a collimated wind with the surrounding medium. The radio string is located
within a massive dense molecular core, and is associated with extended green emission
(Spitzer 3-color), Herbig-Haro type emission (2MASS K$_s$-band) and OH maser sites -- all
phenomena readily observed towards sites of massive star formation.  We conclude that the
massive core hosts a high-mass star in an early stage of evolution in which it is
undergoing the ejection of a powerful collimated stellar wind, showing that jets found in
the formation of low-mass stars are also produced in high-mass stars.

\end{abstract}  

\keywords{ISM: (IRAS 16562$-$3959) --- ISM: jets and outflows
 --- radio continuum: stars --- stars: formation }

\vfill\eject

\section{INTRODUCTION}

The determination whether massive stars are formed via accretion or via merging
processes is one of the main observational challenges in the field of star formation. If
massive O-type stars are formed by an accretion process similar to that inferred for low-mass
stars, then we expect that circumstellar disks and jets will be present in their earliest
stages of evolution. 

To date there are only a handful of massive young stellar objects (YSO) known to be
associated with highly collimated jets and/or Herbig-Haro (HH) objects. All except one
have luminosities smaller than $2\times10^4$ \Lsun\ corresponding to that of a B0 ZAMS
star. They include IRAS 18162-2048 ($L\sim1.7\times10^4$ \Lsun; \citealp{Marti1993ApJ});
Cepheus A HW2 ($L\sim1\times10^4$ \Lsun; \citealt{Rodriguez1994ApJ}); IRAS 20126+4104
($L\sim1.3\times10^4$ \Lsun; \citealp{Cesaroni1997AA}); G192.16-3.82 ($L\sim3\times10^3$
\Lsun; \citealp{Shepherd1998ApJ}; \citealp{Devine1999AJ}; \citealp{Shepherd2001Sci}); and
W75N, which contains several molecular and HH outflows powered by at least four late to
early-B protostars \citep{Shepherd2003ApJ}. There is only one YSO with $L>2\times10^4$
\Lsun\ that is associated with a highly collimated jet (IRAS 16547-4247,
$L\sim6.2\times10^4$ \Lsun; \citealp{Garay2003ApJ}; \citealp{Rodriguez2005ApJ};
\citealp{Brooks2007ApJ}; \citealp{Rodriguez2008AJ}). Two other luminous YSOs, IRAS
18089-1732 ($L\sim 3.2\times10^4$ \Lsun; \citealp{Beuther2008ApJ}) and G331.51$-$0.10
($L\sim1\times10^5$ \Lsun; \citealp{Bronfman2008ApJ}), are associated with radio continuum
sources with spectral indices characteristic of collimated stellar winds, but the angular
resolution of the observations is insufficient to resolve the jet/flow morphology. It is
not clear whether the lack of young massive stars with spectral types earlier than B0 ZAMS
associated with jets and/or disks is an intrinsic property of the most massive stars or
due to observational disadvantages -- massive stars are rarer and their evolutionary time
scales are much shorter than those of low mass stars.

Currently we are carrying out a systematic search for jets towards massive YSOs to assess
whether or not they are a common phenomena (Guzm\'an 2011, in preparation).  In this paper
we report the discovery, made using the Australia Telescope Compact Array (ATCA), of a
string of radio continuum emission associated with IRAS 16562$-$3959. Assuming that this
source is located at the distance of $1.6$ kpc \citep[V$_{\rm LSR}~=-12.6$ km
  s$^{-1}$,][]{Urquhart2008AA}\footnote{The two-fold distance ambiguity was resolved in
  \citet{Faundez2004AA}}, the total far-infrared luminosity, computed using the observed
IRAS fluxes \citep[see][]{Casoli1986AA}, is $\sim7.0\times10^4$ \Lsun. The target was
selected from a list of sources compiled by us with IRAS luminosities in excess of
$2\times10^4$ \Lsun\ and radio emission much weaker than that expected from the total FIR
luminosity. The expectation is that objects with these characteristics are high-mass
objects in the pre-UC \ion{H}{2} region phase, in which the weak radio emission is most likely
arising from stellar wind phenomena \citep[see Fig. 6,][]{Hoare2007prpl}.  IRAS
16562$-$3959 was observed as part of the RMS Survey \citep[G345.4938+01.4677,
][]{Urquhart2008ASPC}.  The emission detected at 4.8 and 8.6 GHz with ATCA
\citep{Urquhart2007AA} was weaker than expected. Moreover, the positive radio spectral
index between these two frequencies and the hint of a string of multiple sources gave us
strong indications that this source could be an ionized jet associated with a massive
young stellar object (MYSO).

In this paper we present new ATCA data taken at 4 frequencies (1.4, 2.4, 4.8 and 8.6 GHz).
Based on these data, we propose that the string of five radio sources detected towards
IRAS 16562$-$3959 is composed of a central thermal jet, plus inner and outer lobes of
shock ionized gas resulting from the interaction of the highly collimated stellar wind
with the surrounding medium.

\section{OBSERVATIONS}

The radio continuum observations were made using the Australia Telescope Compact Array
(ATCA)
\footnote{The Australia Telescope Compact Array is funded by the Commonwealth of Australia
  for operation as a National Facility managed by CSIRO} during 2008 June, October and
2009 February. We used the 1.5B, 1.5C and 6.0A configurations, utilizing all six antennas
and covering east-west baselines from \mbox{30 m} to 5.9 km. Observations were made at
four frequencies: 1.384, 2.368, 4.800 and 8.640 GHz, each with a bandwidth of 128 MHz,
full Stokes. Throughout this work we will refer to these frequencies as  
1.4, 2.4, 4.8 and 8.6 GHz, respectively.  The phase center of the array was
RA$\,=$\ 16$^h$59$^m$41.61$^s$,\ DEC$\,=-40^\circ03\arcmin43.4\arcsec$ (J2000). The total
integration time at each frequency was about 180 minutes, obtained from 10-minute scans
taken over a wide range of hour angles to provide good (u,v) coverage.  The calibrator PKS
1740-517 was observed for 3 min. before and after every on-source scan in order to correct
the amplitude and phase of the interferometer data for atmospheric and instrumental
effects as well as to calibrate the bandpass.

The flux density was calibrated by observing PKS 1934-638 (3C84) for which values of
14.95, 11.59, 5.83, and 2.84 Jy were adopted at 1.4, 2.4, 4.8, and 8.6 GHz,
respectively. Standard calibration and data reduction were performed using MIRIAD
\citep{Sault1995ASPC}. Maps were made by Fourier transformation of the uniformly weighted
interferometer data. The noise level achieved in the images are 0.38, 0.21, 0.096, and
0.070 mJy beam$^{-1}$ and the  synthesized (FWHM) beams obtained were
$10.23\arcsec\times5.71\arcsec$, $5.97\arcsec\times3.27\arcsec$,
$2.81\arcsec\times1.83\arcsec$, and $1.62\arcsec\times1.01\arcsec$, at the frequencies of
1.4, 2.4, 4.8, and 8.6 GHz, respectively.  We estimate the flux uncertainty to be
approximately 20\%.

{\section{RESULTS}\label{section-results}}

\begin{figure}
\plotone{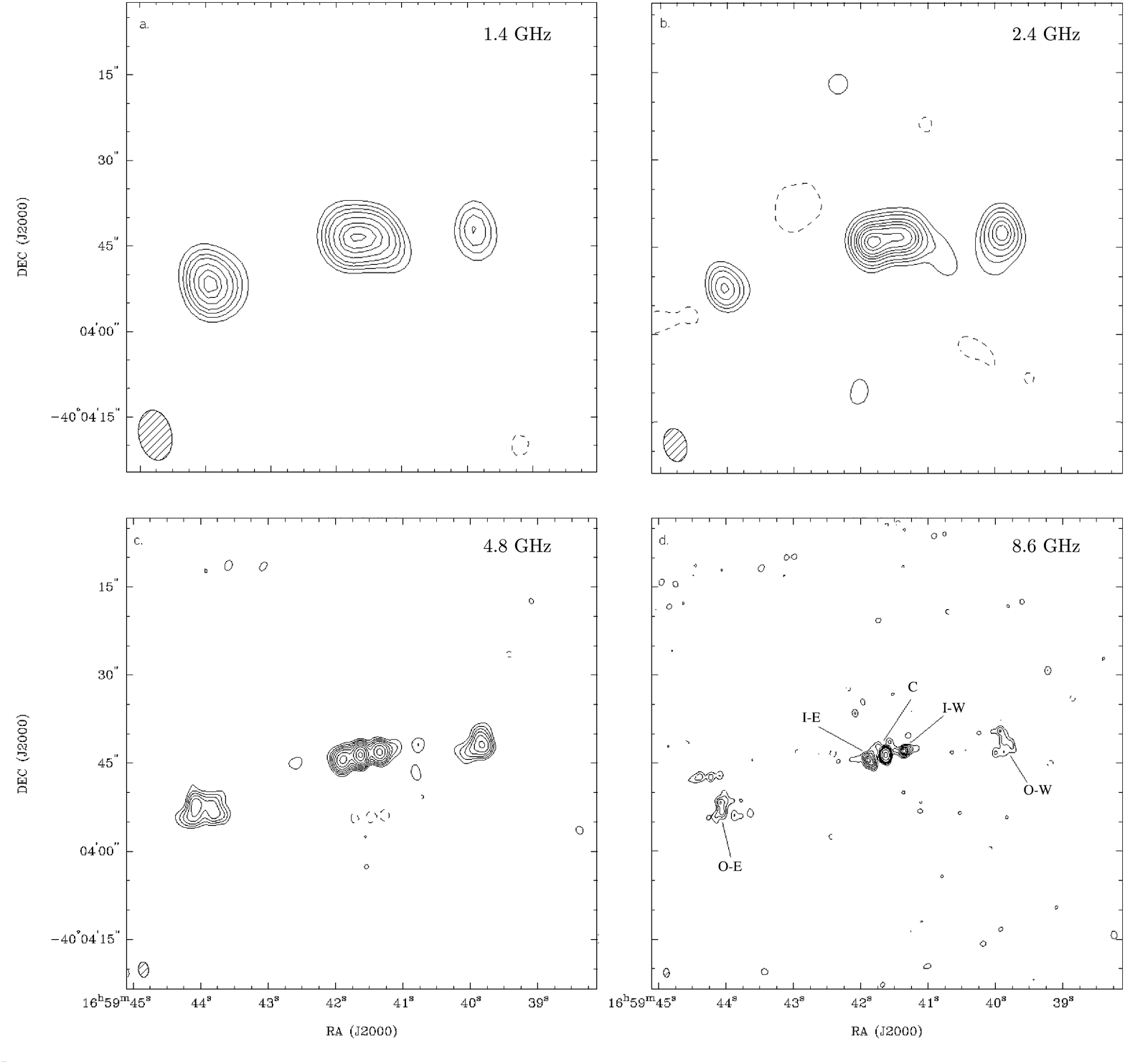}
\figcaption
{\baselineskip0.0pt
ATCA maps of the radio continuum emission from IRAS 16562-3959.
Beams are shown in the lower left corner of each panel.
Top left: 1.4 GHz map. Contour levels are -4, 4, 6, 8, 10, 13, 16, 18, and 20 times
$\sigma$ ($1\sigma$ = 0.35 mJy beam$^{-1}$).
Top right: 2.4 GHz map. Contour levels are -7, 7, 12, 17, 25, 32, 38, 45, 54, and 60 times 
$\sigma$  ($1\sigma$ = 0.12 mJy beam$^{-1}$).
Bottom left: 4.8 GHz map. Contour levels are -4, 3, 6, 9, 13, 19, 28, 40, 60, and 80 times 
$\sigma$ ($1\sigma$ = 0.086 mJy beam$^{-1}$).
Bottom right: 8.6 GHz map. Contour levels are -4,  3,  6,  9,  13, 16, 19, 24, 50, and 100 times $\sigma$  ($1\sigma$ = 0.081 mJy beam$^{-1}$).
\label{fig-radiomap}}
\end{figure}

Figure~\ref{fig-radiomap} shows maps of the radio continuum emission from IRAS
16562$-$3959 at the four frequencies observed with ATCA. The maps at the lower 
frequencies (1.4 and 2.4 GHz) show three sources roughly lying in a linear 
structure along a direction with P.A. of 99.1\arcdeg.  The outermost sources 
which we will call Outer-East and Outer-West, labeled O-E and O-W in 
Fig.~\ref{fig-radiomap}, respectively, are separated by an angular distance of 
$\sim45$\arcsec\ (corresponding to 0.35 pc at 1.6 kpc).  The maps at the higher 
frequencies (4.8 and 8.6 GHz) show that the central object seen at low
frequencies is resolved into three components aligned along a direction with 
P.A. $98.9^\circ$. The external components of this triplet, which we will call 
Inner-East and Inner-West components (labeled I-E and I-W) are symmetrically 
located in opposite directions from the bright central source (labeled C), 
and separated by $\approx7$\arcsec.  Throughout this work, we will use
the abbreviations C, I-E, I-W, O-E, and O-W for the Central, Inner-East, Inner-West,
Outer-East, and Outer-West components, respectively.

We note that the five radio components are not exactly aligned, 
exhibiting a small bending. The lines joining the east components and the central 
source and the west components and the central source form an angle of 168$^\circ$.
We also note that in the 8.6 GHz image there are three small knots located about 
4\arcsec\ north from the Outer-East lobe, with peak fluxes of 0.6 mJy. The knots are well 
aligned with the Inner and Central components suggesting that they could be part 
of a string of emission, but more sensitive observations are needed to confirm this 
or disprove as artifacts of the data reduction.  We will not discuss these 
components further in this work.

The position and flux densities of all radio components detected towards 
IRAS 16562$-$3959 are given in Table \ref{tbl-obs}. For all sources the flux 
densities were determined from Gaussian fittings using the MIRIAD task IMFIT, 
with the exception of the outer components at 4.8 and 8.6 GHz. For these, the 
flux density was measured over the 3$\sigma$ contour level. To estimate the 
flux densities of the C, Inner-East and Inner-West components at 1.4 and 2.4 GHz 
we fitted the observed morphology with three unresolved Gaussians centered at 
the positions determined from the 8.6 GHz observations.

\begin{figure}
  \includegraphics[angle=-90]{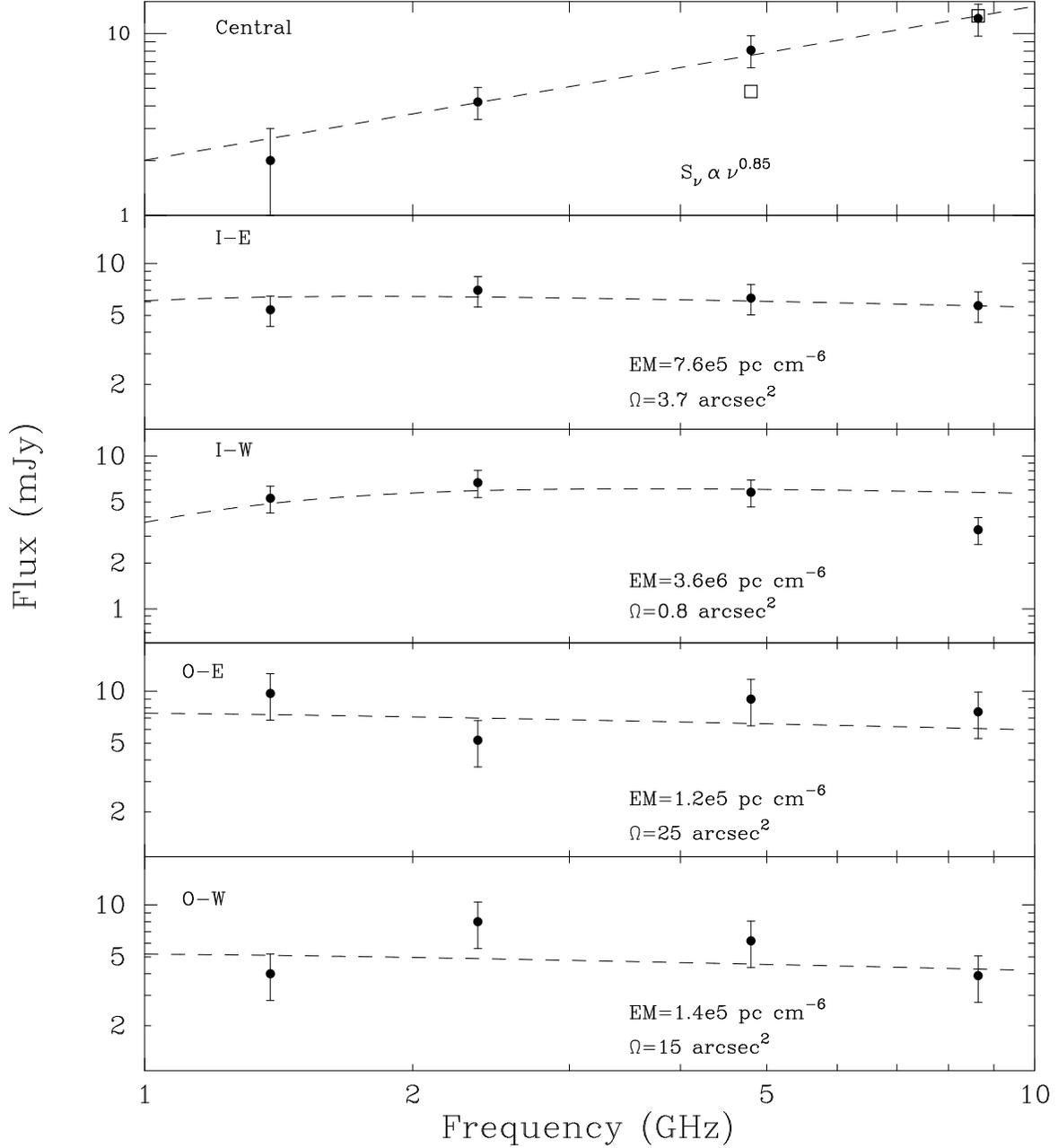} \figcaption{Radio continuum flux density versus
    frequency for the five principal radio components detected toward IRAS
    16562-3959. From top to bottom, spectra of the Central, Inner-East, Inner-West,
    Outer-East and Outer-West components. In the top panel the dashed line indicates a
    least squares power law fit to the data.  Empty squares represent the data reported by
    \citet{Urquhart2007AA}.  In the other panels the dashed line indicates a least square
    fits to the observed spectra with a model of homogeneous region of ionized gas.  The
    emission measure (EM) was fitted, and the solid angle ($\Omega$) was obtained from the
    deconvolved size of the radio images at 8.6 GHz. Both parameters are given in each of the
    panels. Error bars are the maximum of either the error values in Table \ref{tbl-obs}
    or 20\%.
  \label{fig-radioespec}}
\end{figure}

Figure~\ref{fig-radioespec} shows the SED of the integrated radio continuum emission from
the five components. Also plotted in the spectra of the central component are the values
reported by \citet{Urquhart2007AA} at 4.8 and 8.6 GHz of 4.8 and 12.5 mJy,
respectively. We find that the radio continuum spectra of component C is well fitted by a
power law spectrum ($S_{\nu} \propto \nu^{\alpha}$, where $S_{\nu}$ is the flux density at
the frequency $\nu$), with an spectral index $\alpha$ of $0.85\pm0.15$.  For the other
components the data were fitted using a model of thermal emission from a homogeneous
region of ionized gas. The parameters EM and $\Omega$, emission measure and
solid angle, are given in each panel.

\section{DISCUSSION}

\subsection{The nature of the radio sources}

The spectral index of the radio continuum emission from component C, of 
$0.85\pm0.15$, indicates free-free emission arising from a thermal jet. Theoretical
calculations show that collimated stellar winds can have spectral indices in the range
0.25 to 1.1 depending on the radial dependence of the physical quantities of the jet
\citep{Reynolds1986ApJ}.  They also predict that the angular size depends
with frequency.  However, the angular resolution of our observations does not 
allow us to investigate this dependence and observations with higher angular 
resolution are needed to resolve the jet.

Assuming that the Central source is a bipolar, pressure confined jet, 
which has an spectral index of 0.84,  Eqns.(13) and
(19) of \citet{Reynolds1986ApJ} yields, for a distance of 1.6 kpc and an 
observed flux density at 8.6 GHz of 12.1 mJy, the following constraints on the 
jet physical parameters:
\begin{align}
 \left(\frac{n_{{\rm jet}}(r)}{~10^6\,{\rm cm}^{-3}~} \right) &= 
  1.32 \left(\frac{r}{10^{-3}{\rm pc}} \right)^{-0.9} \left( \frac{\nu_m}{10 {\rm GHz}}\right)^{0.818}\left( \frac{\theta}{0.2}
 \right)^{-0.7} \left( \frac{\sin~i}{\sin~45\arcdeg} \right)^{0.3}
\label{reynolds-dens} \\ 
\left(\frac{{\dot M}_w}{10^{-6}~\Msun~yr^{-1}}\right) &= 2.8
\left(\frac{v_w}{10^{3}~ \kms}\right) \left( \frac{\nu_m}{10 {\rm GHz}}\right)^{0.18}
\left( \frac{\theta}{0.2} \right)^{3/4} \left( \frac{\sin~i}{\sin~45\arcdeg}
\right)^{-1/4}
~~, \label{reynolds-accecon}
 \end{align}
where $n_{{\rm jet}}(r)$ is the number density of the jet at a distance $r$ from 
the MYSO, $\nu_m$ is the turnover frequency, $\theta$ the opening angle at 
the base of the jet, $i$ the inclination angle with respect to the line of sight, 
$v_w$ the wind velocity  and ${\dot M}_w$ is the mass loss rate. Most of
 these parameters are unknown. In deriving these expressions, in addition to the 
observational constraint, we assumed an ionization fraction of 1 and a temperature of 
8000 K at the  base of the jet; and a mean particle mass of $1.3~m_H$ per hydrogen 
atom.  

The degree of collimation of the jet source can be estimated from the size of the lobes
and their distance to the central component.  Using the data obtained for the East and
West outer lobes we derive that $\theta \sim0.2$ radians.  Assuming that $v_w$ is $500$
\kms, a value typical of jets associated with luminous objects
\citep{Anglada1996ASPC,Marti1998ApJ,Curiel2006ApJ,Rodriguez2008AJ}, then the constraint
equation \eqref{reynolds-accecon} implies that the mass loss rate is $\sim
1.4\times10^{-6}$ \Msun~yr$^{-1}$.

If the total luminosity of IRAS 16562$-$3959 is produced by a ZAMS star it would
correspond to an O8 star, which emits a rate of ionizing photons of $2.2\times10^{48}$
s$^{-1}$ \citep{Panagia1973AJ}. Embedded in a constant density medium, this star would
 generate an \hii\ region with a flux density of $\sim 8.4$ Jy at optically
thin radio frequencies, far in excess of the observed value of $\sim$ 10 mJy. We suggest
that the weak radio emission from the central source is a consequence of IRAS
16562$-$3959 undergoing an intense accretion phase, with the central object still
being in the pre-UC\hii\ region  sequence phase. 
The high-mass accretion rate of the infalling
material forbids the development of a sizeable \hii\ region
\citep{Yorke1979AA,Walmsley1995RMxAC}, and the free-free emission from the ionized
material is considerably lowered at centimeter wavelengths.

The spectra of the radio emission from the lobes are rather flat, indicative of 
optically thin free-free emission. We suggest that the radio emission from the 
lobes arises in shocks resulting from the interaction of a collimated stellar 
wind with the surrounding medium. The dotted lines in Fig. \ref{fig-radioespec} 
correspond to fits of the observed spectra using a model of thermal emission 
from a homogeneous region of ionized gas. From these fits we derive
emission measures of $\sim1\times10^6$ pc~cm$^{-6}$ towards the inner lobes and
$\sim1\times10^5$ pc~cm$^{-6}$ towards the outer lobes. For all lobes the 
emission is optically thin within the range of observed frequencies. The solid 
angles were derived from the radio images giving 3.7, 0.8, 25 and 15 arcsec$^2$ 
for the Inner-East, Inner-West, Outer-East and Outer-West lobes, respectively. 
The derived electron densities within the lobes are about 
$\sim10^3$ cm$^{-3}$ for the outer lobes and $\sim10^4$ cm$^{-3}$ for the 
inner lobes.

\subsection{Characteristics of the exciting source and its environment}
The radio string is associated with a massive ($\sim1.0\times10^3$ \Msun) and dense
($1.2\times10^6$ cm$^{-3}$) core detected at 250 GHz (1.2 mm) by \citet{Faundez2004AA},
with the Central component offset from the millimeter continuum peak by $\sim$13\arcsec,
equivalent to 0.1 pc at 1.6 kpc.

\begin{figure}
  \includegraphics[angle=-90]{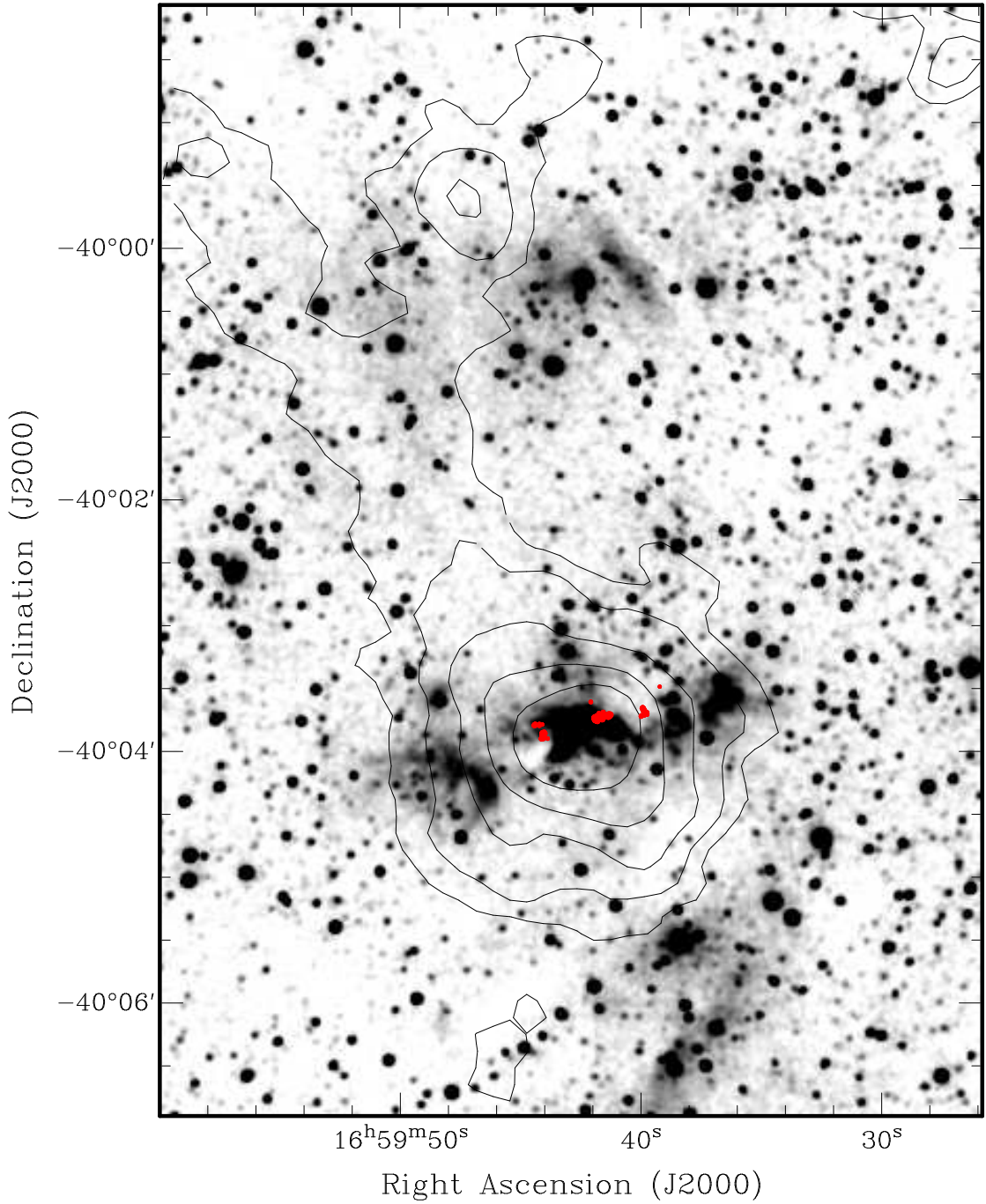}
  \figcaption{Gray scale image of the K$_s$ band emission towards IRAS 16562-3959. 
Black contours: 1.2 mm dust continuum emission (SIMBA/SEST).
Red contours: 8.6 GHz radio continuum emission (ATCA).
  \label{fig-2mass-simba}}
\end{figure}

Figure \ref{fig-2mass-simba} shows that associated with the dust core there is 
strong diffuse K$_s$ band emission, centered near the Central source and 
aligned along the jet axis. This emission extends $\sim1.5$\arcmin\ in the East 
and West directions.  There is also diffuse K$_s$ band emission 
arising from two regions located $\sim3$\arcmin\ towards the North and South
of the Central source. They are roughly symmetrically displaced from the Central 
source but aligned in a perpendicular direction to the jet axis.  
The strong K$_s$ band emission is likely to be produced by 
excited H$_2$ 2.12 $\um$ emission arising from shocked gas. Moreover, the
arc-like morphology of the northern K$_s$ band emission feature located 
approximately at RA$\,=$16$^h$59$^m$42$^s$, DEC$=-40^\circ00\arcmin13\arcsec$ 
(J2000) is also consistent with HH phenomena. We note that the North-South 
structure is also seen as diffuse mid-infrared emission in the four MSX 
bands (8.28, 12.13,  14.65, and 21.34~\um), indicating also 
a strong contribution from warm dust.

\begin{figure}
\plotone{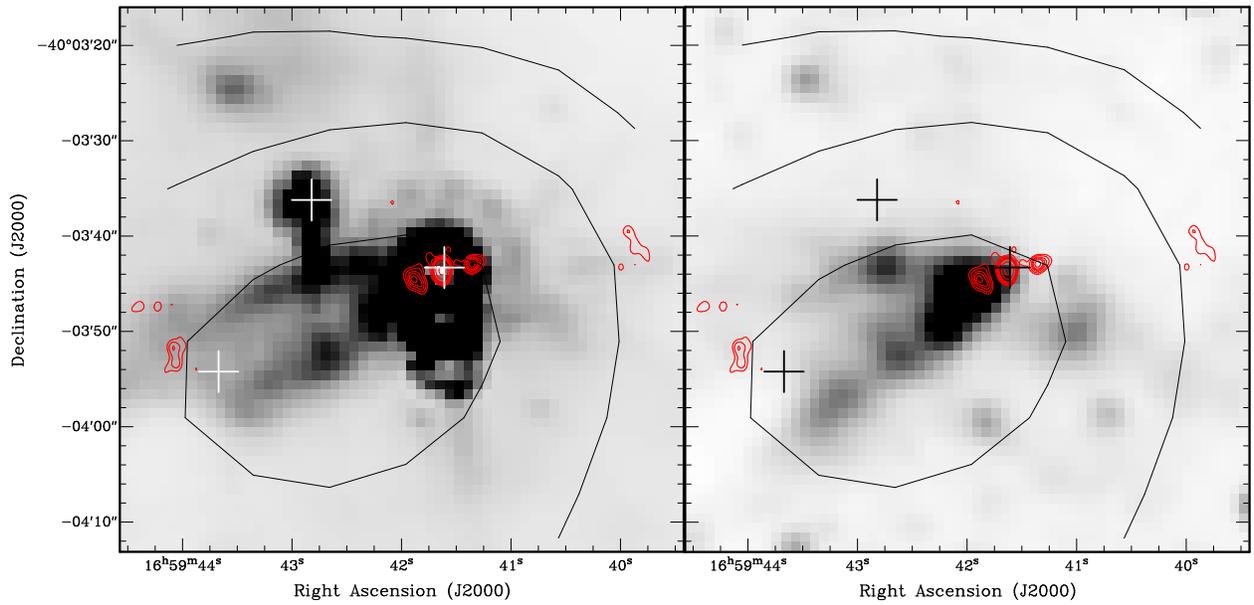}
  \figcaption{{\it Left}: Gray image: 8 $\um$ IRAC-Spitzer emission. 
    Black contours: 250 GHz dust emission detected with SIMBA.
Red contours: 8.6 GHz emission. Crosses mark the position of the OH masers. 
Note that the 8 $\um$ data is saturated near the central source, 
which produces an spurious extension running towards the south. 
{\it Right}: Gray image: K$_s$ band emission. Contours and symbols same  
as in the left panel.
  \label{fig-2massand8um}}
\end{figure}

\begin{figure}
  \plotone{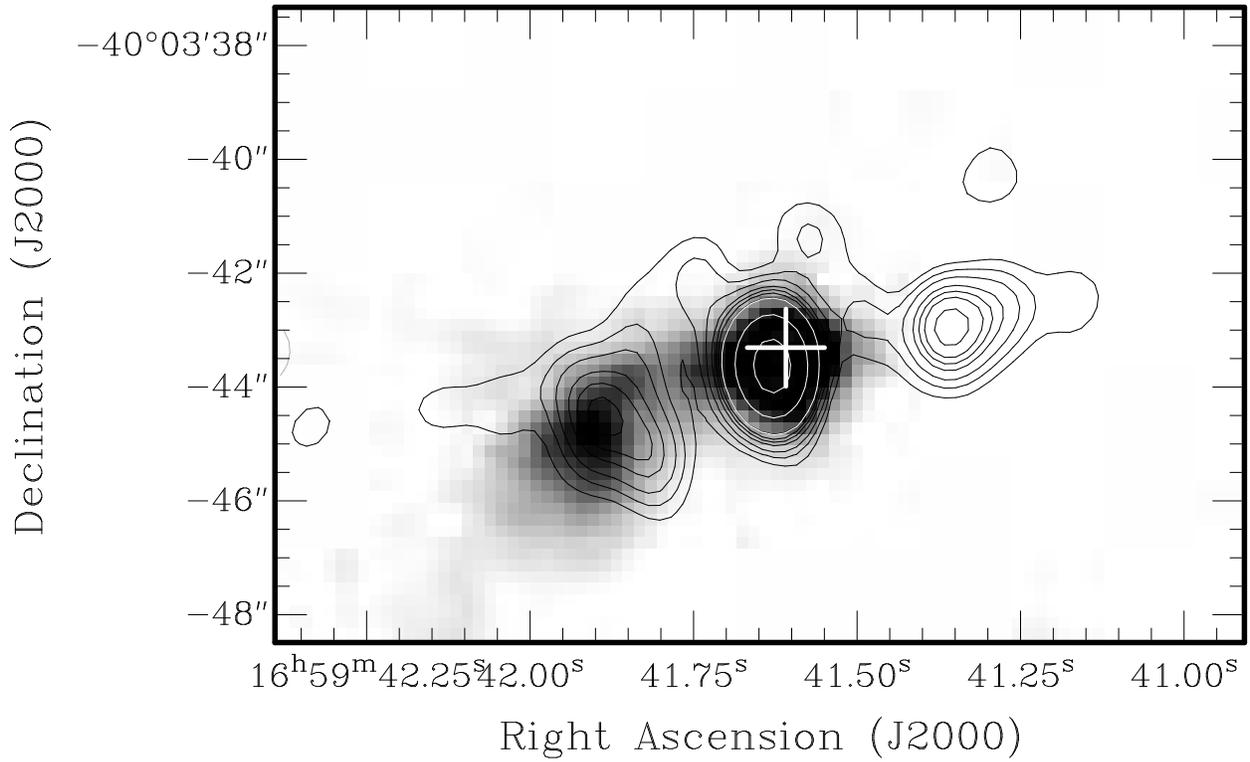}
  \figcaption{ {\it Grey scale}: TIMMI2 10.4 $\um$ emission. {\it Contours}: 8.6 GHz radio
    emission.  The cross marks the position of the OH maser associated with the
    central source. There is strong 10.4 $\um$ emission associated with the central source
    and more diffuse emission associated with the Inner-East lobe.
  \label{fig-timmi2central}}
\end{figure}

\begin{figure}
  \plotone{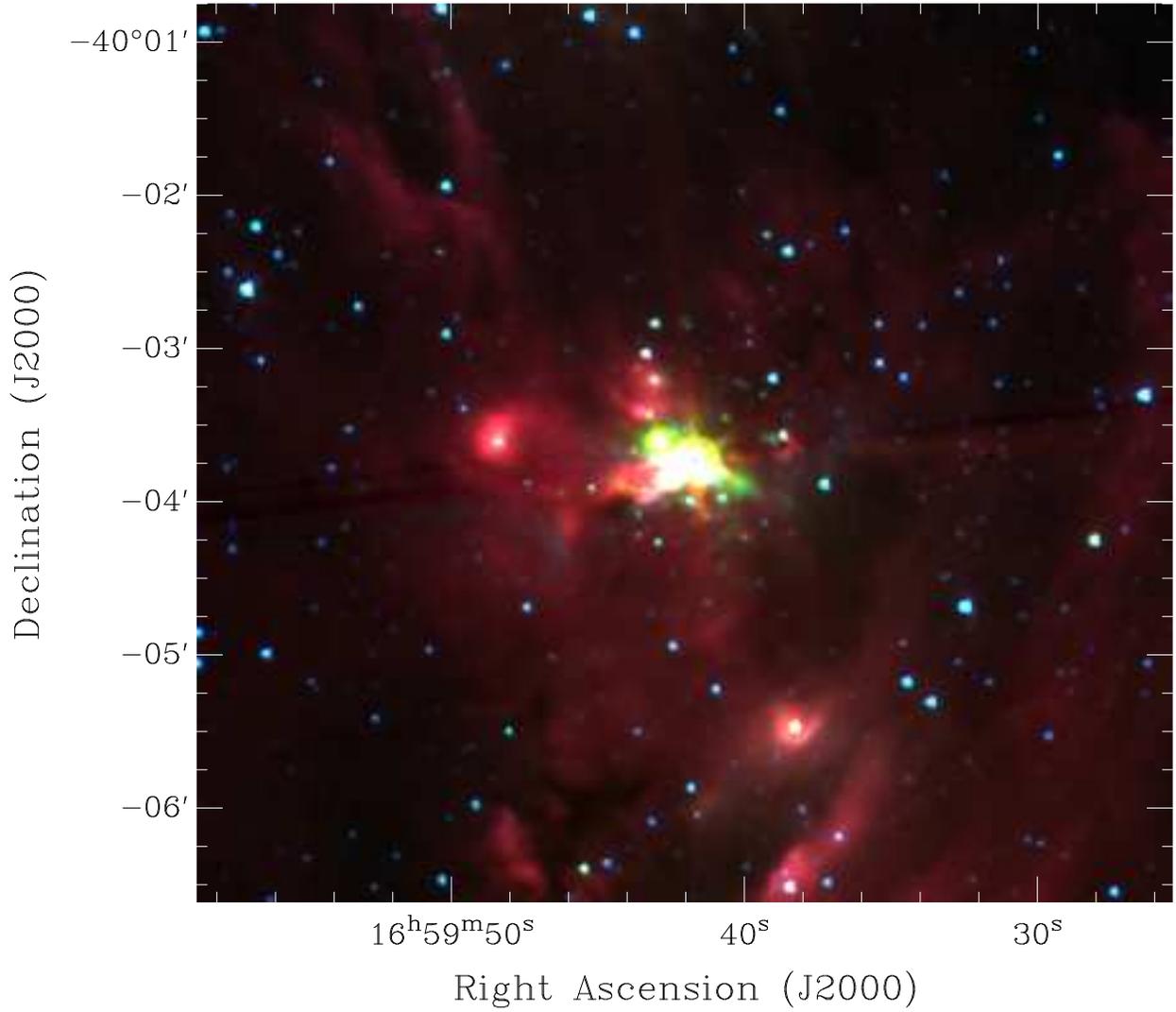}
  \figcaption{ Spitzer three color image using the 8.0 \um,
   4.5 \um and 3.6 \um IRAC images for red, green and blue,
    respectively. The jet is located at the center of the image associated with 
 an extended diffuse ``green fuzzy'' or enhancement of the 4.5 \um band.  
  \label{fig-Spitzer3color}}
\end{figure}

Figure \ref{fig-2massand8um} shows a zoom in of the 8 $\um$ emission from 
Spitzer and K$_s$ band emission from 2MASS towards the central part of the core.  
There is an extended ``V'' shape feature extending along the East jet axis seen 
in both K$_s$ band and 8 \um. In addition, there are three OH masers 
\citep{Caswell1998MNRAS297215, Caswell2004MNRAS}: one is associated with the 
central source, a second is associated with a 8 \um source, unseen in 2MASS, 
located $\sim15\arcsec$ northeast of the central source and the third is located 
close to the East jet axis at about 2\arcsec\ from the Outer-East lobe.  Their 
velocities are close to the radial velocity of the ambient cloud
\citep[V$_{\rm LSR}~=-12.6$ \kms,][]{Urquhart2008AA}, except for the eastern OH 
maser, which has a radial velocity of $-24.5$ \kms.  Infrared continuum emission 
at 10.4 \um is shown in
Fig. \ref{fig-timmi2central}, which displays an image\footnote{Downloaded from
  http://www.ast.leeds.ac.uk/RMS/} obtained with TIMMI2 by \citet{Mottram2007AA}. It shows
intense emission associated with the Central source and more diffuse emission associated
with the Inner-East radio lobe.  There is also an extended 4.5 \um emission enhancement
associated to the central source (see Fig. \ref{fig-Spitzer3color}). These ``green
fuzzies'', as seen green in a three color IRAC images, have been related to shocked gas in
protostellar environments \citep{Chambers2009APJS}.  All of the phenomena described above
is readily associated with high-mass star formation and therefore support the notion that
IRAS 16562-3959 is a high-mass young stellar object with energetic outflow activity.

While the East-West K$_s$ band structure can be related directly to the 
radio jet, the identification of the energy source of the North-South K$_s$ band
feature is less clear. One possibility is that the North-South feature is driven 
by the Central jet source, which will imply a re-orientation of the jet in almost 
90$^\circ$. Models of close stellar encounters do not, however, predict such 
large precession of the outflow axis \citep[e.g.][]{Moeckel2006ApJ}, and thus 
this possibility appears unlikely.  Since massive YSOs are known to be formed 
in clusters, the most plausible explanation is that a yet unidentified nearby 
source is the exciting source of the North-South K$_s$-band emission.
A possible candidate is the YSO, located 16\arcsec\ north-east of the Central component,  
identified from the Spitzer data. As pointed out previously, this object is 
associated to an OH maser with a radial velocity similar to that of the ambient 
cloud. This and the Central source belong thus to the
same star forming region, as suggested by \citet{Caswell1998MNRAS297215}.  

\begin{figure}
  \plotone{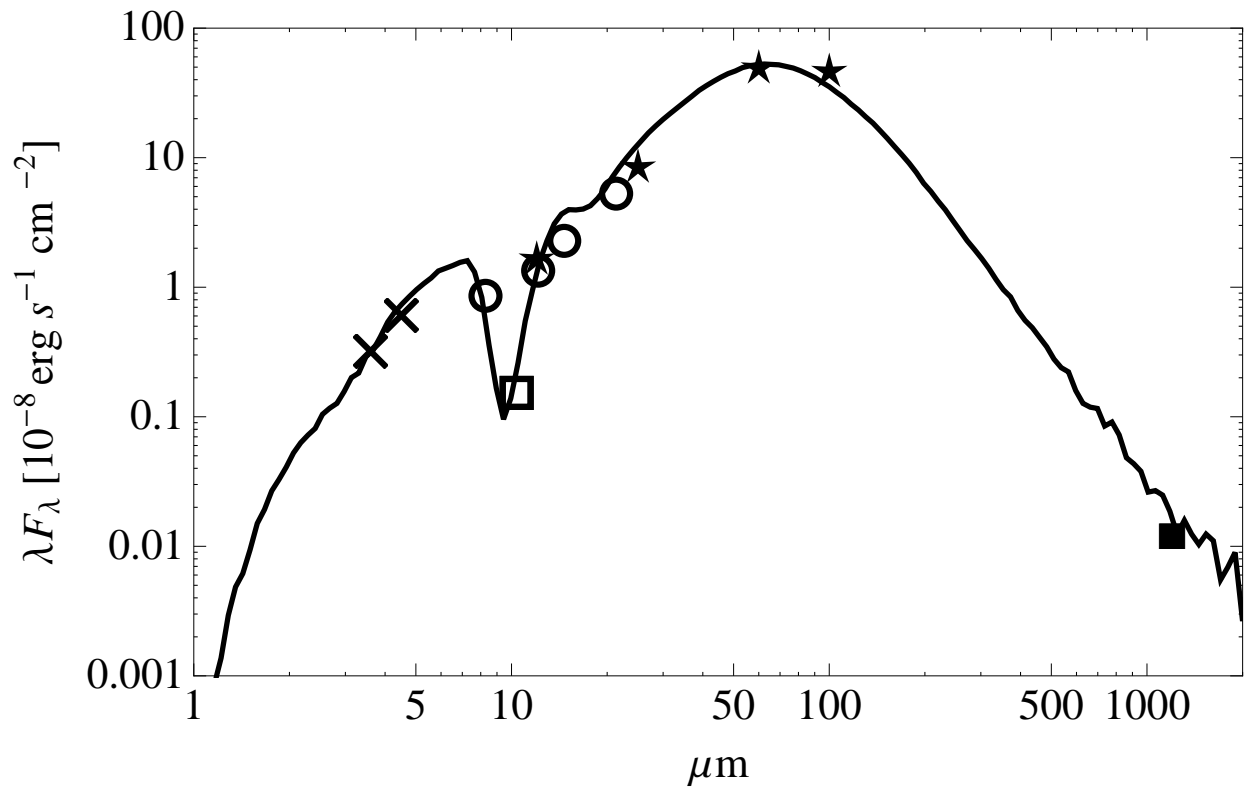}
  \figcaption{Spectral energy distribution of the Central source. 
{\it Crosses}: Spitzer/IRAC data at 3.6 and 4.5 $\mu$m. 
{\it Empty square}: TIMMI2 data at 10.4 $\mu$m. 
{\it Empty circles}: MSX data at 8.3, 12.1, 14.7 and 21.3 $\mu$m. 
{\it Stars}: IRAS data at 12, 25, 60 and 100 $\um$. 
{\it Filled Square}: SIMBA data at 1200 $\um$.  
The line indicates the best fit using the models described in 
\citet{Robitaille2007ApJS}. \label{fig-sedRob}}
\end{figure}

Figure~\ref{fig-sedRob} shows the spectral energy distribution (SED) of 
IRAS 16562-3959 from 3.6 \um to 1200 \um, including flux densities at 
1.2~millimeter \citep{Faundez2004AA}; at 12, 25, 60, and 100~\micron\ obtained 
from the IRAS database; at
8.3, 12.1, 14.7, and 21.3~\micron\ obtained from the \emph{Midcourse Space Experiment
  (MSX)} Survey of the Galactic Plane database \citep[MSX source
  G345.4938+01.4677]{Price2001AJ}, and at 3.6 and 4.5 \um obtained from Spitzer-IRAC
images using an aperture of $\sim1'$. The IRAC images at 5.8 and 8.0 \um data were
saturated and an adequate flux estimation was not possible. In the frequency range covered
by the SED the emission is mainly due to thermal dust emission.  The spectral energy
distribution was analyzed by fitting model SEDs using a large grid of precomputed models
\citep{Robitaille2007ApJS}.  The continuous line in Fig.~\ref{fig-sedRob} presents the
result of the best fit.  The derived parameters from the fit are given in Tab.
\ref{tbl-robfit}.  The SED fitting indicates that the energy source corresponds to a
deeply embedded YSO, and rules out the possibility of being a heavily extincted star.

The mass of the host envelope derived from the fitting is $\sim1700$ M$_\odot$. However,
the model of \citet{Robitaille2007ApJS} does not consider the increase of absorption at mm
frequencies due to ice coatings in the dust grains \citep{Ossenkopf1994AA} and in
consequence is probably overestimating the mass of the envelope by a factor $2\sim3$.
Therefore, the mass of the core derived from the 1.2 mm observations of 910
M$_\odot$ is in fair agreement with the SED fitting. We note that \citet{Faundez2004AA}
reported a mass of $1500$ M$_\odot$, but a new analysis of the data shows that for this
particular source an erroneous calibration factor was applied, which produced an
overestimation of the flux by a factor of 1.76 (L\'opez et al. 2010 in preparation).

The fitting also indicates that the central YSO has a mass of $\sim15~ $\Msun\ and a
luminosity of $5.8\times10^4$ \Lsun, that the envelope is undergoing an intense accretion
phase with a rate of $3.4\times10^{-3}$ \Msun~yr$^{-1}$ and that the accretion rate onto
the central object is $5.5\times10^{-4}$ \Msun~yr$^{-1}$.  The derived stellar parameters
are similar to those of a B0$\sim$09.5 main sequence star \citep{Sternberg2003ApJ} and a
O8.5 ZAMS star \citep{Panagia1973AJ}. We also note that the derived infall accretion rate
is greater than $\sim 1.0\times 10^{-5}$ \Msun~yr$^{-1}$, the infall rate needed to quench
the development of a sizeable \ion{H}{2} region \citep[]{Walmsley1995RMxAC}.

The linear morphology and spectral characteristics of the radio sources suggest that the
string is physically associated with a highly collimated wind arising from a young,
massive YSO with the lobes tracing the interaction between the jet and the ambient cloud.
We postulate that the lobes radio emission corresponds to thermal emission coming from
shocked ionized gas situated in the working surfaces of the Central jet source.
Particularly, we rule out the possibility that the radio emission has an extragalactic
origin -- the probability of finding an extragalactic source at 5 GHz with a flux density
above 5 mJy within $20\arcsec$\ from the peak of the 1.2 mm dust emission is
$3.5\times10^{-4}$ \citep{Fomalont1991AJ}.

\subsection{Jet and Shock parameters}

In this section we make use of the observed quantities of the lobe emission
to derive parameters of the jet-ambient medium interaction. 
Considering that the free-free emission from the lobes arises from a shock 
wave formed at the head of the jet, then the emission measure, EM, is given 
in terms of shock parameters by \citep{Curiel1993ApJ}
\begin{equation}
 \left({\frac{{\rm EM}}{10^{6}~{\rm pc~cm}^{-6}}}\right)
 = 1.39 \left({\frac{n_a}{10^{5}~ {\rm cm}^{-3}}}\right)
\left({\frac{V_s}{100~ {\rm km~s}^{-1}}}\right)^{1.68}
\left({\frac{T_e}{10^4 ~{\rm K}}}\right)^{0.8}~~,
\label{curiel-shock}
\end{equation}
where $n_a$, $V_s$ and $T_e$ are the particle ambient density, the shock velocity 
and the temperature of the shock-ionized gas. If the ambient density is known,
this expression can be used to estimate the shock velocity.

To estimate the ambient density at the position of the 
lobes we re-analyzed the 1.2-mm data taken by \citeauthor{Faundez2004AA} 
toward IRAS 16562$-$3959 and fitted the observed radial intensity profiles 
from the core region using a model in which the density and temperature follow 
power-law radial distributions
\citep{Adams1991ApJ}.  
The best fit to the image was obtained with the following molecular gas density 
and temperature profiles: 
$n_{a}(r)= n_0(0.1\text{ pc}/r)^{1.9}$ and 
$T(r) =T_0 (0.1\text{ pc}/r)^{0.4}$, 
 where $n_0=1.4\times 10^{5}$ cm$^{-3}$ and $T_0=50$ K.
We assumed a gas to dust mass ratio of 100 and a mean molecular
mass of 2.3 $m_H$. The derived density profile implies that the mass of the core, 
within a radius of $\sim90\arcsec$, is $790$ \Msun,
and that the ambient density at the position of the East and West
inner lobes -- displaced from the core center by 10\arcsec\ and 18\arcsec --
are $2.3\times10^5$ and $7.6\times10^4$ cm$^{-3}$, respectively.
Using these densities and the emission
measures computed in \S \ref{section-results}, we derive from
Eq. \eqref{curiel-shock}\ that the shock velocities at the East and West inner 
lobes are 119 and 91 \kms, respectively.  The same analysis applied to the East 
and West outer lobes gives shock velocities of $\sim$24 and $\sim$66 km s$^{-1}$, 
respectively, consistent with the expectation that shock velocity should decrease 
away from the jet source.

Assuming a simple one dimensional isothermal shock model 
\citep[e.g.][]{Masson1993ApJ} and that the momentum flux imparted to the shock 
ionized material by the jet equals the ram pressure of the ambient material, then
\begin{equation}
\label{jet-vel} 
\left( V_j-V_s \right)^2\rho_j=V_s^2\rho_a\,,  
\end{equation}
where $\rho_j$ and $V_j$ are the mass density and velocity of the jet, 
respectively,  and $\rho_a$ is the ambient gas mass density.
If the shock velocity, ambient density and jet density are known this expression 
can be used to estimate the jet velocity.
The density of the jet at the positions of the inner lobes can be estimated from
Eq. \eqref{reynolds-dens}.  The lobes are displaced from the jet by $3\arcsec$, 
which translates into a physical distance of 0.033 pc, assuming an inclination 
of $45^\circ$.  Using $\theta$=0.2 and assuming $\nu_m$= 10 GHz, then the density 
of the jet at 0.033 pc is  $1.2\times 10^{-19}$ gr cm$^{-3}$, 
about $\sim 4$ times lower than the ambient density.
Using the values derived for the East and West inner lobes of the 
shock velocity, ambient density and jet density we
determine from Eq. \eqref{jet-vel} jet velocities of 360 and 280 km s$^{-1}$, 
respectively.

We further estimate that the most recent ejection, giving rise to the inner 
lobes, took place between 90 and 120 years ago.  Moreover, from Eq. 
\eqref{reynolds-accecon} and taking again 10 GHz as the turnover frequency, we 
derive mass loss rates of 1.0 and 0.79 $\times10^{-6}$\Msun~yr$^{-1}$, 
which in combination with the ejecta velocities give momentum rates of 
$3.7\times10^{-4}$ and $2.2\times10^{-4}$ \Msun\ km s$^{-1}$ yr$^{-1}$ for the 
East and West halves of the bipolar jet, respectively. 

\section{SUMMARY}

We made radio continuum observations at 1.4, 2.4, 4.8 and 8.6 GHz, using ATCA, toward IRAS
16562-3959, a luminous object ($L\sim7\times10^4$\ \Lsun) thought to be a massive star
forming region in an early stage of evolution.  The main results and conclusions are
summarized as follows:
\begin{enumerate}
\item{ The radio continuum observations show the presence of a remarkable string of radio
  emission, consisting of a compact, bright central component, two inner lobes, separated
  by about 7\arcsec\ and symmetrically located from the central source, and two outer
  lobes, separated by about 45\arcsec.}

\item{ The emission from the central object has a spectral index between 1.4 and 8.6 GHz
  of $0.85\pm0.15$, consistent with free-free emission from a thermal jet.  Assuming that
  the jet corresponds to a bipolar pressure confined wind with an aperture angle
  $\theta=0.2$, we estimate that the jet has a total mass loss rate of $\sim 2\times10^{-6}$  \Msun~yr$^{-1}$.
}

\item{ The radio emission from the lobes have spectral indices of typically $-0.1$,
  characteristic of thermal emission. We suggest that the emission from the lobes arises
  in shocks resulting from the interaction of a collimated wind with the surrounding
  medium. }

\item{ The string is found projected towards a massive and dense core and associated with
several indicators of massive star formation and shock gas tracers. The jet is 
  located near the peak position of the dust emission.  We conclude that the massive core
  hosts a high-mass star in an early stage of evolution in which it is undergoing the
  ejection of a powerful collimated stellar wind, showing that jets found in the formation
  of low-mass stars are also produced in high-mass stars.}
\end{enumerate}

\acknowledgments

The authors gratefully acknowledge support from CONICYT through projects FONDAP
No. 15010003 and BASAL PFB-06. This publication made use of the GLIMPSE-Spitzer database.
  This paper also made use of information from the Red MSX Source
survey database at www.ast.leeds.ac.uk/RMS which was constructed with support from the
Science and Technology Facilities Council of the UK.

\newpage



\begin{deluxetable}{cllccccc}
  \tablewidth{0pt}
  \tablecaption{OBSERVED PARAMETERS OF RADIO SOURCES \label{tbl-obs}}
  \tablehead{
    \colhead{Source}      & \multicolumn{2}{c}{8.6 GHz peak position}  &  &
    \multicolumn{4}{c}{Flux density (mJy)}  \\
    \cline{2-3} \cline{5-8}
    \colhead{}  & \colhead{$\alpha$(2000)}  & \colhead{$\delta$(2000)} & \colhead{~}
    & \colhead{1.4 GHz} & \colhead{2.4 GHz} & \colhead{4.8 GHz} & \colhead{8.6 GHz} \\
  }
  \startdata
  Central (C) & $16^{\rm h} 59^{\rm m} 41{\rlap.}{^s}63$ & $-40\arcdeg\ 03\arcmin\
  43{\rlap.}{\arcsec}61$  & ~
  
  & $2.0\pm1.0$ & $4.2\pm0.8$ & $8.1\pm0.2$ & $12.1\pm0.2$ \\
  
  Inner-East (I-E) & 16~59~~41.87 & $-40~~03~~44.55$ & ~
  & $5.4\pm0.9$ & $7.0\pm0.7$ & $6.3\pm0.2$ & $5.7\pm0.1$ \\ 
  
  Inner-West (I-W) & 16~59~~41.35 &  $-40~~03~~42.94$ & ~
  & $5.3\pm0.9$ & $6.7\pm0.7$ & $5.8\pm0.2$ & $3.3\pm0.2$ \\ 
  
  Outer-East (O-E)&16~59~~44.07 & $-40~~03~~52.21$ & ~
  & $9.7\pm0.3$ & $5.2\pm0.2$ & $9.0\pm 0.3 $ & $7.6\pm 0.4$ \\ 
  
  Outer-West (O-W)& 16~59~~39.83 &  $-40~~03~~41.90$ & ~
  & $4.0\pm 0.7$ & $8.0\pm 0.5 $ & $6.2\pm 0.2 $ & $3.9\pm 0.3$ \\

  \enddata
\end{deluxetable}

\begin{deluxetable}{rl}
  \tablewidth{0pt}
  \tablecaption{SED FITTING \label{tbl-robfit}}
  \tablehead{
    \colhead{Parameter}      & \colhead{Value}
  }
  \startdata
  Inclination & 41$^\circ$\\
  Stellar Age & $4.5\times10^4$ yr\\
  Stellar Mass & $14.7$ \Msun \\
  Envelope accretion rate & $3.4\times10^{-3}$ \Msun~yr$^{-1}$\\
  Disk Mass & $0.26$ \Msun\\
  Disk accretion rate & $5.5\times10^{-4}$ \Msun~yr$^{-1}$\\
  Total Luminosity & $5.8\times10^4$ \Lsun\\
  Envelope Mass &$1.7\times10^3$ \Msun\\
  \enddata
\end{deluxetable}





\clearpage 

\bibliographystyle{apj}
\bibliography{bibliografia}

\begin{thebibliography}{40}
\expandafter\ifx\csname natexlab\endcsname\relax\def\natexlab#1{#1}\fi

\bibitem[{{Adams}(1991)}]{Adams1991ApJ}
{Adams}, F.~C. 1991, \apj, 382, 544

\bibitem[{{Anglada}(1996)}]{Anglada1996ASPC}
{Anglada}, G. 1996, in Astronomical Society of the Pacific Conference Series,
  Vol.~93, Radio Emission from the Stars and the Sun, ed. {A.~R.~Taylor \&
  J.~M.~Paredes}, 3--7

\bibitem[{{Beuther} \& {Walsh}(2008)}]{Beuther2008ApJ}
{Beuther}, H., \& {Walsh}, A.~J. 2008, \apjl, 673, L55

\bibitem[{{Bronfman} {et~al.}(2008){Bronfman}, {Garay}, {Merello}, {Mardones},
  {May}, {Brooks}, {Nyman}, \& {G{\"u}sten}}]{Bronfman2008ApJ}
{Bronfman}, L., {Garay}, G., {Merello}, M., {Mardones}, D., {May}, J.,
  {Brooks}, K.~J., {Nyman}, L.-{\AA}., \& {G{\"u}sten}, R. 2008, \apj, 672, 391

\bibitem[{{Brooks} {et~al.}(2007){Brooks}, {Garay}, {Voronkov}, \&
  {Rodr{\'{\i}}guez}}]{Brooks2007ApJ}
{Brooks}, K.~J., {Garay}, G., {Voronkov}, M., \& {Rodr{\'{\i}}guez}, L.~F.
  2007, \apj, 669, 459

\bibitem[{{Casoli} {et~al.}(1986){Casoli}, {Combes}, {Dupraz}, {Gerin}, \&
  {Boulanger}}]{Casoli1986AA}
{Casoli}, F., {Combes}, F., {Dupraz}, C., {Gerin}, M., \& {Boulanger}, F. 1986,
  \aap, 169, 281

\bibitem[{{Caswell}(1998)}]{Caswell1998MNRAS297215}
{Caswell}, J.~L. 1998, \mnras, 297, 215

\bibitem[{{Caswell}(2004)}]{Caswell2004MNRAS}
---. 2004, \mnras, 349, 99

\bibitem[{{Cesaroni} {et~al.}(1997){Cesaroni}, {Felli}, {Testi}, {Walmsley}, \&
  {Olmi}}]{Cesaroni1997AA}
{Cesaroni}, R., {Felli}, M., {Testi}, L., {Walmsley}, C.~M., \& {Olmi}, L.
  1997, \aap, 325, 725

\bibitem[{{Chambers} {et~al.}(2009){Chambers}, {Jackson}, {Rathborne}, \&
  {Simon}}]{Chambers2009APJS}
{Chambers}, E.~T., {Jackson}, J.~M., {Rathborne}, J.~M., \& {Simon}, R. 2009,
  \apjs, 181, 360

\bibitem[{{Curiel} {et~al.}(1993){Curiel}, {Rodriguez}, {Moran}, \&
  {Canto}}]{Curiel1993ApJ}
{Curiel}, S., {Rodriguez}, L.~F., {Moran}, J.~M., \& {Canto}, J. 1993, \apj,
  415, 191

\bibitem[{{Curiel} {et~al.}(2006){Curiel}, {Ho}, {Patel}, {Torrelles},
  {Rodr{\'{\i}}guez}, {Trinidad}, {Cant{\'o}}, {Hern{\'a}ndez}, {G{\'o}mez},
  {Garay}, \& {Anglada}}]{Curiel2006ApJ}
{Curiel}, S., {et~al.} 2006, \apj, 638, 878

\bibitem[{{Devine} {et~al.}(1999){Devine}, {Bally}, {Reipurth}, {Shepherd}, \&
  {Watson}}]{Devine1999AJ}
{Devine}, D., {Bally}, J., {Reipurth}, B., {Shepherd}, D., \& {Watson}, A.
  1999, \aj, 117, 2919

\bibitem[{{Fa{\'u}ndez} {et~al.}(2004){Fa{\'u}ndez}, {Bronfman}, {Garay},
  {Chini}, {Nyman}, \& {May}}]{Faundez2004AA}
{Fa{\'u}ndez}, S., {Bronfman}, L., {Garay}, G., {Chini}, R., {Nyman},
  L.-{\AA}., \& {May}, J. 2004, \aap, 426, 97

\bibitem[{{Fomalont} {et~al.}(1991){Fomalont}, {Windhorst}, {Kristian}, \&
  {Kellerman}}]{Fomalont1991AJ}
{Fomalont}, E.~B., {Windhorst}, R.~A., {Kristian}, J.~A., \& {Kellerman}, K.~I.
  1991, \aj, 102, 1258

\bibitem[{{Garay} {et~al.}(2003){Garay}, {Brooks}, {Mardones}, \&
  {Norris}}]{Garay2003ApJ}
{Garay}, G., {Brooks}, K.~J., {Mardones}, D., \& {Norris}, R.~P. 2003, \apj,
  587, 739

\bibitem[{{Hoare} {et~al.}(2007){Hoare}, {Kurtz}, {Lizano}, {Keto}, \&
  {Hofner}}]{Hoare2007prpl}
{Hoare}, M.~G., {Kurtz}, S.~E., {Lizano}, S., {Keto}, E., \& {Hofner}, P. 2007,
  Protostars and Planets V, 181

\bibitem[{{Mart\'i} {et~al.}(1993){Mart\'i}, {Rodr\'iguez}, \&
  {Reipurth}}]{Marti1993ApJ}
{Mart\'i}, J., {Rodr\'iguez}, L.~F., \& {Reipurth}, B. 1993, \apj, 416, 208

\bibitem[{{Mart\'i} {et~al.}(1998){Mart\'i}, {Rodr\'iguez}, \&
  {Reipurth}}]{Marti1998ApJ}
---. 1998, \apj, 502, 337

\bibitem[{{Masson} \& {Chernin}(1993)}]{Masson1993ApJ}
{Masson}, C.~R., \& {Chernin}, L.~M. 1993, \apj, 414, 230

\bibitem[{{Moeckel} \& {Bally}(2006)}]{Moeckel2006ApJ}
{Moeckel}, N., \& {Bally}, J. 2006, \apj, 653, 437

\bibitem[{{Mottram} {et~al.}(2007){Mottram}, {Hoare}, {Lumsden}, {Oudmaijer},
  {Urquhart}, {Sheret}, {Clarke}, \& {Allsopp}}]{Mottram2007AA}
{Mottram}, J.~C., {Hoare}, M.~G., {Lumsden}, S.~L., {Oudmaijer}, R.~D.,
  {Urquhart}, J.~S., {Sheret}, T.~L., {Clarke}, A.~J., \& {Allsopp}, J. 2007,
  \aap, 476, 1019

\bibitem[{{Ossenkopf} \& {Henning}(1994)}]{Ossenkopf1994AA}
{Ossenkopf}, V., \& {Henning}, T. 1994, \aap, 291, 943

\bibitem[{{Panagia}(1973)}]{Panagia1973AJ}
{Panagia}, N. 1973, \aj, 78, 929

\bibitem[{{Price} {et~al.}(2001){Price}, {Egan}, {Carey}, {Mizuno}, \&
  {Kuchar}}]{Price2001AJ}
{Price}, S.~D., {Egan}, M.~P., {Carey}, S.~J., {Mizuno}, D.~R., \& {Kuchar},
  T.~A. 2001, \aj, 121, 2819

\bibitem[{{Reynolds}(1986)}]{Reynolds1986ApJ}
{Reynolds}, S.~P. 1986, \apj, 304, 713

\bibitem[{{Robitaille} {et~al.}(2007){Robitaille}, {Whitney}, {Indebetouw}, \&
  {Wood}}]{Robitaille2007ApJS}
{Robitaille}, T.~P., {Whitney}, B.~A., {Indebetouw}, R., \& {Wood}, K. 2007,
  \apjs, 169, 328

\bibitem[{{Rodr{\'{\i}}guez} {et~al.}(2005){Rodr{\'{\i}}guez}, {Garay},
  {Brooks}, \& {Mardones}}]{Rodriguez2005ApJ}
{Rodr{\'{\i}}guez}, L.~F., {Garay}, G., {Brooks}, K.~J., \& {Mardones}, D.
  2005, \apj, 626, 953

\bibitem[{{Rodr\'iguez} {et~al.}(1994){Rodr\'iguez}, {Garay}, {Curiel},
  {Ram\'irez}, {Torrelles}, {G\'omez}, \& {Velazquez}}]{Rodriguez1994ApJ}
{Rodr\'iguez}, L.~F., {Garay}, G., {Curiel}, S., {Ram\'irez}, S., {Torrelles},
  J.~M., {G\'omez}, Y., \& {Velazquez}, A. 1994, \apjl, 430, L65

\bibitem[{{Rodr{\'{\i}}guez} {et~al.}(2008){Rodr{\'{\i}}guez}, {Moran},
  {Franco-Hern{\'a}ndez}, {Garay}, {Brooks}, \& {Mardones}}]{Rodriguez2008AJ}
{Rodr{\'{\i}}guez}, L.~F., {Moran}, J.~M., {Franco-Hern{\'a}ndez}, R., {Garay},
  G., {Brooks}, K.~J., \& {Mardones}, D. 2008, \aj, 135, 2370

\bibitem[{{Sault} {et~al.}(1995){Sault}, {Teuben}, \& {Wright}}]{Sault1995ASPC}
{Sault}, R.~J., {Teuben}, P.~J., \& {Wright}, M.~C.~H. 1995, in Astronomical
  Society of the Pacific Conference Series, Vol.~77, Astronomical Data Analysis
  Software and Systems IV, ed. {R.~A.~Shaw, H.~E.~Payne, \& J.~J.~E.~Hayes},
  433

\bibitem[{{Shepherd} {et~al.}(2001){Shepherd}, {Claussen}, \&
  {Kurtz}}]{Shepherd2001Sci}
{Shepherd}, D.~S., {Claussen}, M.~J., \& {Kurtz}, S.~E. 2001, Science, 292,
  1513

\bibitem[{{Shepherd} {et~al.}(2003){Shepherd}, {Testi}, \&
  {Stark}}]{Shepherd2003ApJ}
{Shepherd}, D.~S., {Testi}, L., \& {Stark}, D.~P. 2003, \apj, 584, 882

\bibitem[{{Shepherd} {et~al.}(1998){Shepherd}, {Watson}, {Sargent}, \&
  {Churchwell}}]{Shepherd1998ApJ}
{Shepherd}, D.~S., {Watson}, A.~M., {Sargent}, A.~I., \& {Churchwell}, E. 1998,
  \apj, 507, 861

\bibitem[{{Sternberg} {et~al.}(2003){Sternberg}, {Hoffmann}, \&
  {Pauldrach}}]{Sternberg2003ApJ}
{Sternberg}, A., {Hoffmann}, T.~L., \& {Pauldrach}, A.~W.~A. 2003, \apj, 599,
  1333

\bibitem[{{Urquhart} {et~al.}(2007){Urquhart}, {Busfield}, {Hoare}, {Lumsden},
  {Clarke}, {Moore}, {Mottram}, \& {Oudmaijer}}]{Urquhart2007AA}
{Urquhart}, J.~S., {Busfield}, A.~L., {Hoare}, M.~G., {Lumsden}, S.~L.,
  {Clarke}, A.~J., {Moore}, T.~J.~T., {Mottram}, J.~C., \& {Oudmaijer}, R.~D.
  2007, \aap, 461, 11

\bibitem[{{Urquhart} {et~al.}(2008{\natexlab{a}}){Urquhart}, {Hoare},
  {Lumsden}, {Oudmaijer}, \& {Moore}}]{Urquhart2008ASPC}
{Urquhart}, J.~S., {Hoare}, M.~G., {Lumsden}, S.~L., {Oudmaijer}, R.~D., \&
  {Moore}, T.~J.~T. 2008{\natexlab{a}}, in Astronomical Society of the Pacific
  Conference Series, Vol. 387, Massive Star Formation: Observations Confront
  Theory, ed. {H.~Beuther, H.~Linz, \& T.~Henning}, 381

\bibitem[{{Urquhart} {et~al.}(2008{\natexlab{b}}){Urquhart}, {Busfield},
  {Hoare}, {Lumsden}, {Oudmaijer}, {Moore}, {Gibb}, {Purcell}, {Burton},
  {Mar{\'e}chal}, {Jiang}, \& {Wang}}]{Urquhart2008AA}
{Urquhart}, J.~S., {et~al.} 2008{\natexlab{b}}, \aap, 487, 253

\bibitem[{{Walmsley}(1995)}]{Walmsley1995RMxAC}
{Walmsley}, M. 1995, in Revista Mexicana de Astronomia y Astrofisica Conference
  Series, Vol.~1, Revista Mexicana de Astronomia y Astrofisica Conference
  Series, ed. {S.~Lizano \& J.~M.~Torrelles}, 137

\bibitem[{{Yorke}(1979)}]{Yorke1979AA}
{Yorke}, H.~W. 1979, \aap, 80, 308

\end{thebibliography}


\end{document}